\documentclass[journal]{IEEEtran}
\ifCLASSINFOpdf
\usepackage[pdftex]{graphicx}
\graphicspath{figures/}
\DeclareGraphicsExtensions{.pdf,.jpeg,.png}
\else
\usepackage[dvips]{graphicx}
\graphicspath{{figures/}}
\DeclareGraphicsExtensions{.eps}
\fi
\usepackage[cmex10]{amsmath}
\usepackage{bm}
\usepackage{hyperref}
\usepackage{mathtools}
\usepackage{helvet}
\usepackage{xcolor}
\usepackage{cite}
\usepackage{amsfonts}

\usepackage{url}
\begin{document}
\title{All-Spin Bayesian Neural Networks}

\author{Kezhou~Yang, Akul~Malhotra, Sen~Lu,
and~Abhronil~Sengupta,~\IEEEmembership{Member,~IEEE}
\thanks{Manuscript received November, 2019.}
\thanks{All authors contributed equally to this work. The authors are with the School
of Electrical Engineering and Computer Science, Department of Materials Science and Engineering, The Pennsylvania State University, University Park,
PA 16802, USA. A. Malhotra is also affiliated with Birla Institute of Technology and Science, Pilani, Rajasthan 333031, India. E-mail: sengupta@psu.edu.}}
\maketitle
\begin{abstract}
\small{Probabilistic machine learning enabled by the Bayesian formulation has recently gained significant attention in the domain of automated reasoning and decision-making. While impressive strides have been made recently to scale up the performance of deep Bayesian neural networks, they have been primarily standalone software efforts without any regard to the underlying hardware implementation. In this paper, we propose an ``All-Spin" Bayesian Neural Network where the underlying spintronic hardware provides a better match to the Bayesian computing models. To the best of our knowledge, this is the first exploration of a Bayesian neural hardware accelerator enabled by emerging post-CMOS technologies. We develop an experimentally calibrated device-circuit-algorithm co-simulation framework and demonstrate $24\times$ reduction in energy consumption against an iso-network CMOS baseline implementation. }
\end{abstract}


\section{Introduction}
Probabilistic inference is at the core of decision-making in the brain. While the past few years have witnessed unprecedented success of deep learning in a plethora of pattern recognition tasks (complemented by advancements in dedicated hardware designs for these workloads), these problem spaces are usually characterized by the availability of large amounts of data and networks that do not explicitly represent any uncertainty in the network structure or parameters. However, as we strive to deploy Artificial Intelligence platforms in autonomous systems like self-driving cars, decision-making based on uncertainty is crucial. Standard supervised backpropagation based learning techniques are unable to deal with such issues since they do not overtly represent uncertainty in the modelling process. To circumvent these problems, Bayesian deep learning has recently been gaining attention where deep neural networks are trained in a probabilistic framework following the classic rules of probability, i.e. Bayes' theorem. In the Bayesian formulation, the network is visualized as a set of plausible models (assuming \textit{prior} probability distributions on its parameters, for instance, synaptic weights). Given observed data, the \textit{posterior} probability distributions are learnt that best explains the observed data. The key distinction between standard deep networks and Bayesian deep networks is the fact that network parameters in the latter case are modelled as probability distributions. It is worth noting here that the probability distributions are usually modelled by Gaussian processes characterized by a mean and variance \cite{gal2016uncertainty}. Utilizing probability distributions to model network parameters allows us to characterize the network outputs by an uncertainty measure (variance of the distribution), instead of just point estimates in a standard network. These uncertainty measures can therefore be used by autonomous agents for decision making and self-assessment in the presence of continuous streaming data.

This paper explores a hardware-software co-design approach to accelerate Bayesian deep learning platforms through the usage of spintronic technologies. Recent research has demonstrated the possibility of mimicking the primitives of standard deep learning frameworks -- synapses and neurons by single magnetic device structures that can be operated at very low terminal voltages \cite{sengupta2017encoding,sengupta2018stochastic,romera2018vowel}. Further, being non-volatile in nature, spintronic devices can be arranged in crossbar architectures to realize ``In-Memory" dot-product computing kernels -- thereby alleviating the memory access and memory leakage bottlenecks prevalent in CMOS-based implementations \cite{sengupta2016proposal,sengupta2017performance}. As mentioned earlier, the key distinction between Bayesian and standard deep learning is the requirement of sampling from probability distributions and inference based on sampled values. Interestingly, scaled nanomagnetic devices operating at room temperature are characterized by thermal noise resulting in probabilistic switching. We propose to leverage the inherent device stochasticity of spintronic devices to generate samples from Gaussian probability distributions by drawing insights from statistical Central Limit Theorem. Further, our paper also elaborates on a cohesive design of a spintronic Bayesian processor that leverages benefits of spin-based Gaussian random number generators and spintronic ``In-Memory" crossbar architectures to realize high-performance, energy efficient hardware platforms. We believe the drastic reductions in circuit complexity (single devices emulating synaptic scaling operations, crossbar architectures implementing ``In-Memory" dot-product computing kernels and leveraging device stochasticity to sample from probability distributions) and low operating voltages of spintronic devices make them a promising path toward the realization of Probabilistic Machine Learning enabled by the Bayesian formulation.

 \section{Preliminaries: Bayesian Neural Networks}
\begin{figure}[!t]
\centering
\includegraphics[width=2in]{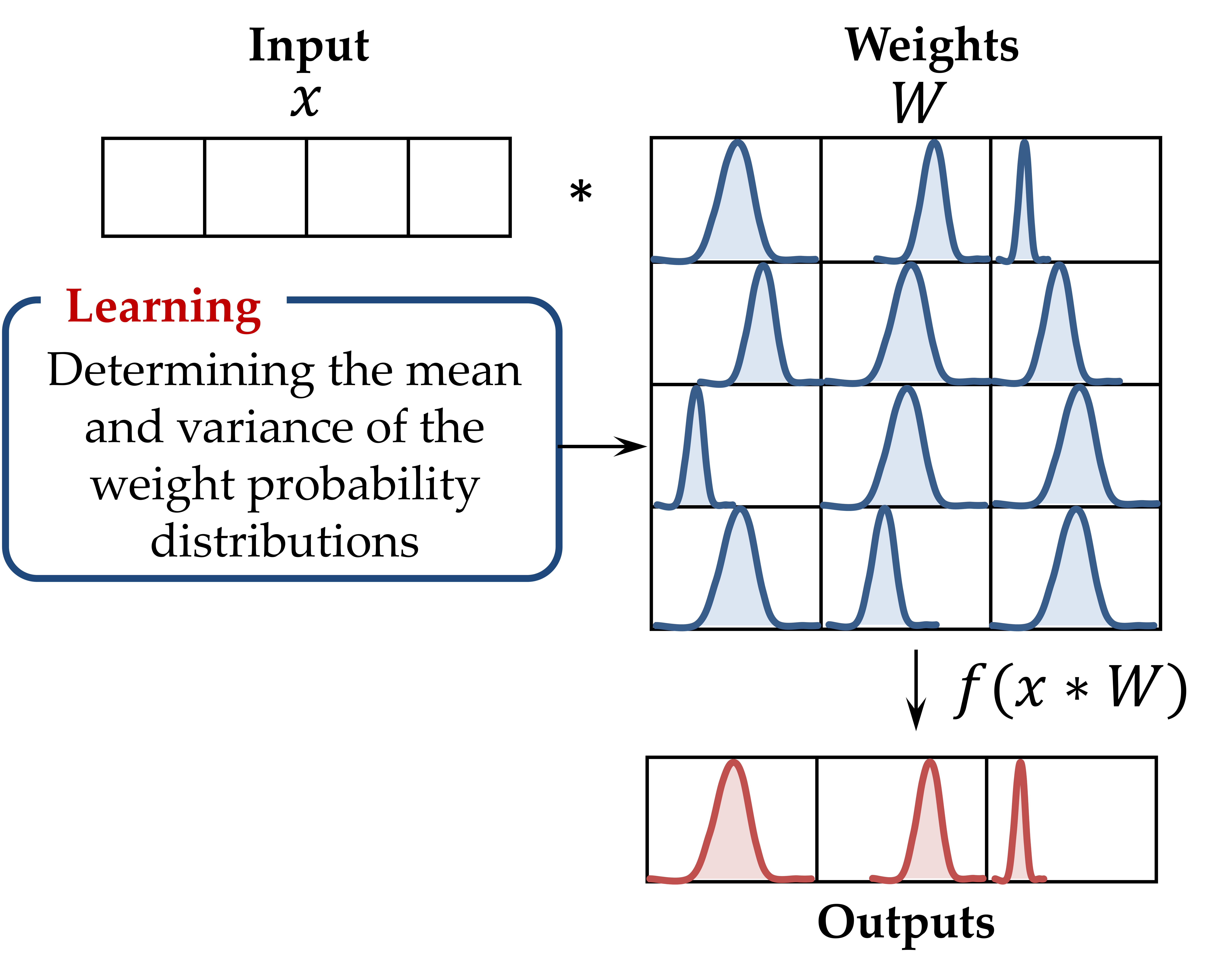}
\caption{In a Bayesian framework, each synaptic weight is represented by a Gaussian probability distribution. The core computing kernel for a particular layer during inference is a dot-product between the inputs and a synaptic weight matrix sample drawn from the individual probability distributions. Learning involves the determination of the mean and variances of the probability distributions using Bayes' formulation.}
\label{fig1}
\end{figure}
Before going into the technical details of the work, we would like to first discuss the preliminaries of Bayesian Neural Networks and the main computationally expensive operations pertaining to their hardware implementation. As shown in Fig. \ref{fig1}, a particular layer of a neural network consists of a set of neurons receiving inputs (sensory information or previous layer of neurons) through synaptic weights, $\textbf{W}$. Bayesian Neural Networks consider the weights of the network, $\textbf{W}$, to be latent variables characterized by a probability distribution, instead of point estimates. More specifically, each weight in such a framework is a random number drawn from a \textit{posterior probability distribution} (characterized by a mean and variance) that is conditioned on a \textit{prior probability distribution} and the \textit{observed datapoints}, $D$ (incoming patterns to the network). Hence, during inference, each incoming data pattern will propagate through the synaptic weights, each of which is characterized by a probability distribution. Hence, as shown in Fig. \ref{fig1}, the final output of the neurons of a particular layer will also be described by a probability distribution characterized by a mean and variance (the uncertainty measure). 

Bayesian Neural Networks correspond to the family of deep learning networks where the weights are `learnt' using Bayes' rule. The learning process here involves the estimation of the mean and variance of the weight \textit{posterior} distribution. Following Bayes' rule, the \textit{posterior probability} can be written as,
\begin{equation}
    P(\textbf{W}|D) = \frac{P(D|\textbf{W})P(\textbf{W})}{P(D)}
\end{equation}
where, $P(\textbf{W})$ denotes the \textit{prior probability} (probability of the latent variables before any data input to the network). $P(D|\textbf{W})$ is the \textit{likelihood}, corresponding to the feedforward pass of the network. In order to make the above \textit{posterior probability} density estimation tractable, two popular approaches are -- Variational Inference methods \cite{houthooft2016vime} or Markov Chain Monte Carlo methods \cite{andrieu2003introduction}. However, in this paper, we focus on Variational Inference methods due to its scalability to large-scale problems \cite{cai2018vibnn}. Variational Inference methods usually approximates the \textit{posterior} distribution by a Gaussian distribution, $q(\textbf{W, $\theta$})$, characterized by parameters, $\theta=(\mu,\sigma$), where $\mu$ and $\sigma$ represent the mean and standard deviation vectors for the probability distributions representing $P(\textbf{W}|D)$ \cite{ghahramani2001propagation}. To summarize, the main hardware design space concerns in Bayesian Neural Networks can be categorized as follows:

\indent {$\bullet$}  \textbf{Gaussian Random Number Generation:} Central to the entire framework, both in the learning as well as the inference process, is the random number generation corresponding to the synaptic weights. Given current large model sizes characterized by over a million synapses, coupled with the fact that random draws need to performed multiple times for each synaptic weight, random number generator circuits would contribute significantly to the total area and power consumption of the hardware. Further, the random numbers need to be sampled from a Gaussian distribution, thereby increasing the complexity of the circuit. We will discuss the hardware costs for CMOS implementations of such Gaussian random number generators in the following sections along with their limitations, followed by our proposal of nanomagnetic random number generators that can serve as the basic building blocks of such Bayesian Neural Networks.
    
\indent {$\bullet$} \textbf{Dot-Product Operation Between Inputs and Sampled Synaptic Weights:} A common aspect of any standard deep learning framework is the fact that forward propagation of information through the network involves a significant amount of memory-intensive operations. The dot-product operation between the synaptic weights and inputs for inference involves the compute energy along with memory access and memory leakage components. For large-scale problems and correspondingly large-scale models, CMOS memory access and memory leakage can be almost $\sim 50\%$ of the total energy consumption profile \cite{ankit2017resparc}. 

The situation is further worsened in a Bayesian deep network since each synaptic weight is characterized by two parameters (mean and variance of the probability distribution), thereby requiring double memory storage. However, the dot-product operation does not occur directly between the inputs and these parameters. In fact, for each inference operation the synaptic weights (typically assumed constant during inference for non-probabilistic networks and implemented by memory elements in hardware) are repeatedly updated depending on sampled values from the Gaussian probability distribution. Hence, direct utilization of crossbar-based ``In-Memory" computing platforms enabled by non-volatile memory technologies (discussed in details later) for alleviating the memory access and memory fetch bottlenecks is not possible and therefore requires a significant rethinking.
   
In the following sections, we sequentially expand on each of these points and propose a spin-based neural processor that merges deterministic and stochastic devices as a potential pathway to enable Bayesian deep learning that can be orders of magnitude more efficient in contrast to state-of-the-art CMOS implementations.

\section{Spintronic Device Design}

\subsection{Magnetic Tunnel Junction - True Random Number Generator Design}

The basic device structure under consideration is the Magnetic Tunnel Junction (MTJ), which consists of two nanomagnets sandwiching a spacer layer (typically an oxide such as MgO). The magnetization of one of the layers is magnetostatically ``pinned" in a particular direction while the magnetization of the other layer can be manipulated by a spin current or an external magnetic field. The two layers are denoted as the ``Pinned" layer (PL) and ``Free" layer (FL). Depending on the relative orientation of the two magnets, the device exhibits a high-resistance anti-parallel (AP) state (when the magnetizations of the two layers have opposite direction) and a low-resistance parallel (P) state (when the magnetizations of the two layers have the same direction). These two states are stabilized by an energy barrier determined by the anisotropy and volume of the magnet.

Let us now consider the switching of the magnet from one state to another by the application of an external current. The switching process is inherently stochastic at non-zero temperatures due to thermal noise \cite{scholz2001micromagnetic}. In the presence of an external current, the probability of switching from one state to the other is modulated depending on the magnitude and duration of the current. True random number generator (TRNG) can be designed using such a device by biasing the magnet at the ``write" current corresponding to a switching probability of 50\%. Note that CMOS-based TRNGs suffer from high energy consumption and circuit design complexity \cite{yang201416}. Proposals and experimental demonstrations of MTJ-based TRNG have been shown \cite{vodenicarevic2017low}. MTJ-based TRNGs are characterized by low area footprint and compatibility with CMOS technology.
 
\begin{figure}
\centering
\includegraphics[width=0.45\textwidth]{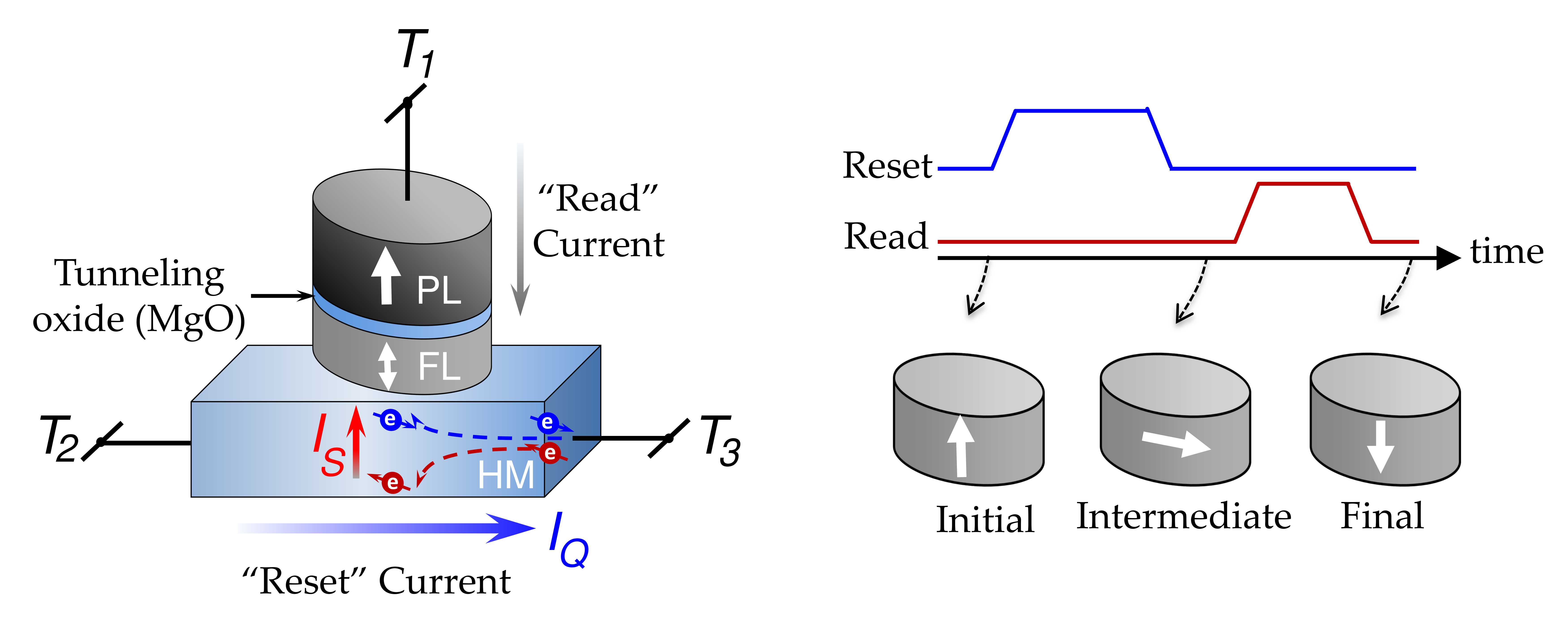}
\caption{The TRNG device structure is shown. Reset current ($I_Q$) flowing through the heavy-metal (HM) results in in-plane spin current ($I_S$) injection for the MTJ ``free layer" (FL). After switching to the in-plane meta-stable position, the magnet relaxes to either of the two stable states with 50\% probability.}
\label{fig2}
\end{figure}

In this paper, we consider a spin-orbit coupling enabled device structure (Fig. \ref{fig2}). It consists of the MTJ stack lying on top of a heavy-metal (HM) underlayer. The device ``read" is performed through the MTJ stack between terminals T1 and T3. However, the device ``write" is performed by passing current through the heavy-metal underlayer between terminals T2 and T3. Input current flowing through the heavy-metal results in spin-injection at the interface of the magnet and heavy-metal due to spin-Hall effect (SHE) \cite{hirsch1999spin} and thereby causes switching of the MTJ ``free layer" \cite{liu2012spin}. The device has the following advantages:

\indent {$\bullet$} The decoupled ``write" and ``read" current paths is advantageous from the perspective of peripheral circuit design to avoid ``read"-``write" conflicts since the associated circuits can be optimized independently. 

\indent {$\bullet$} Such devices offer 1-2 orders of magnitude energy efficiency in comparison to standard spin-transfer torque MRAMs. This is due to the fact that in such spin-orbit coupling based systems, every incoming electron in the ``write" current path repeatedly scatters at the interface of the magnet and heavy metal and transfers multiple units of spin angular momentum to the ferromagnet lying on top.

Usage of SHE-based switching enables us to use an alternative TRNG design \cite{kim2015spin,sengupta2015spin} that has the potential to produce high quality random numbers in presence of process, voltage and temperature (PVT) variations. In the earlier scenario of a standard MTJ, device-to-device variations can result in deviations of the bias current required for 50\% switching probability, thereby degrading the quality of the random number generation process. Our scheme is depicted in Fig. \ref{fig2}, where a magnet with Perpendicular Magnetic Anisotropy (PMA) lies on top of the heavy-metal. The device operation is divided into three stages. During an initial ``Reset" stage, a current flowing through the heavy metal results in in-plane spin injection in the magnet and orients it along the hard-axis for a sufficient magnitude of the ``reset" current. The magnet is then allowed to relax to either of the two stable states in presence of thermal noise -- the switching probability being 50\% since the hard-axis is a meta-stable orientation point for the magnet. In this case, device-to-device variations only causes change in the critical current required for biasing the magnet close to the meta-stable orientation and does not skew the probability distribution to a particular direction (as in the standard MTJ case). Hence, by maintaining a worst-case critical value of the heavy-metal ``reset" current, quality of the random number generation process can be preserved even in the presence of PVT variations. Further, the ``reset" current does not flow through the tunneling oxide layer (unlike the standard MTJ case) and therefore reliability of the oxide layer is not a concern in this scenario \cite{kim2015spin,sengupta2015spin}. Note that our device operation is validated by recent experiments of holding the magnet to its meta-stable hard-axis orientation for performing Bennett clocking in the context of nanomagnetic logic \cite{bhowmik2014spin}. SHE-based energy-efficient switching also results in reduction of the energy consumption involved in the random number generation process. 

The probabilistic switching characteristics of the MTJ can be analyzed by Landau-Lifshitz-Gilbert (LLG) equation with additional term to account for the spin-orbit torque generated by spin-Hall effect at the ferromagnet-heavy metal interface \cite{slonczewski1989conductance},
\begin{equation}
\label{llg}
\frac {d\widehat {\textbf {m}}} {dt} = -\gamma(\widehat {\textbf {m}} \times \textbf {H}_{eff})+ \alpha (\widehat {\textbf {m}} \times \frac {d\widehat {\textbf {m}}} {dt})+\frac{1}{qN_{s}} (\widehat {\textbf {m}} \times \textbf {I}_s \times \widehat {\textbf {m}})
\end{equation}
where, $\widehat {\textbf {m}}$ is the unit vector of FL magnetization, $\gamma= \frac {2 \mu _B \mu_0} {\hbar}$ is the gyromagnetic ratio for electron, $\alpha$ is Gilbert\textquoteright s damping ratio, $\textbf{H}_{eff}$ is the effective magnetic field including the shape anisotropy field for elliptic disks, $N_s=\frac{M_{s}V}{\mu_B}$ is the number of spins in free layer of volume $V$ ($M_{s}$ is saturation magnetization and $\mu_{B}$ is Bohr magneton), and $\textbf{I}_{s}=\theta_{SH}(A_{MTJ}/A_{HM})\textbf{I}_{q}$ is the input spin current ($A_{MTJ}$ and $A_{HM}$ are the MTJ and HM cross-sectional area, $\theta_{SH}$ is the spin-Hall angle and $\textbf{I}_q$ is the charge current flowing through the HM underlayer). Thermal noise is included by an additional thermal field \cite{scholz2001micromagnetic}, $\textbf{H}_{thermal}=\sqrt{\frac{\alpha}{1+\alpha^{2}}\frac{2K_{B}T_{K}}{\gamma\mu_{0}M_{s}V\delta_{t}}}G_{0,1}$, where $G_{0,1}$ is a Gaussian distribution with zero mean and unit standard deviation, $K_{B}$ is Boltzmann constant, $T_{K}$ is the temperature and $\delta_{t}$ is the simulation time-step.
\begin{figure}[t!]
\centering
\includegraphics[width=0.48\textwidth]{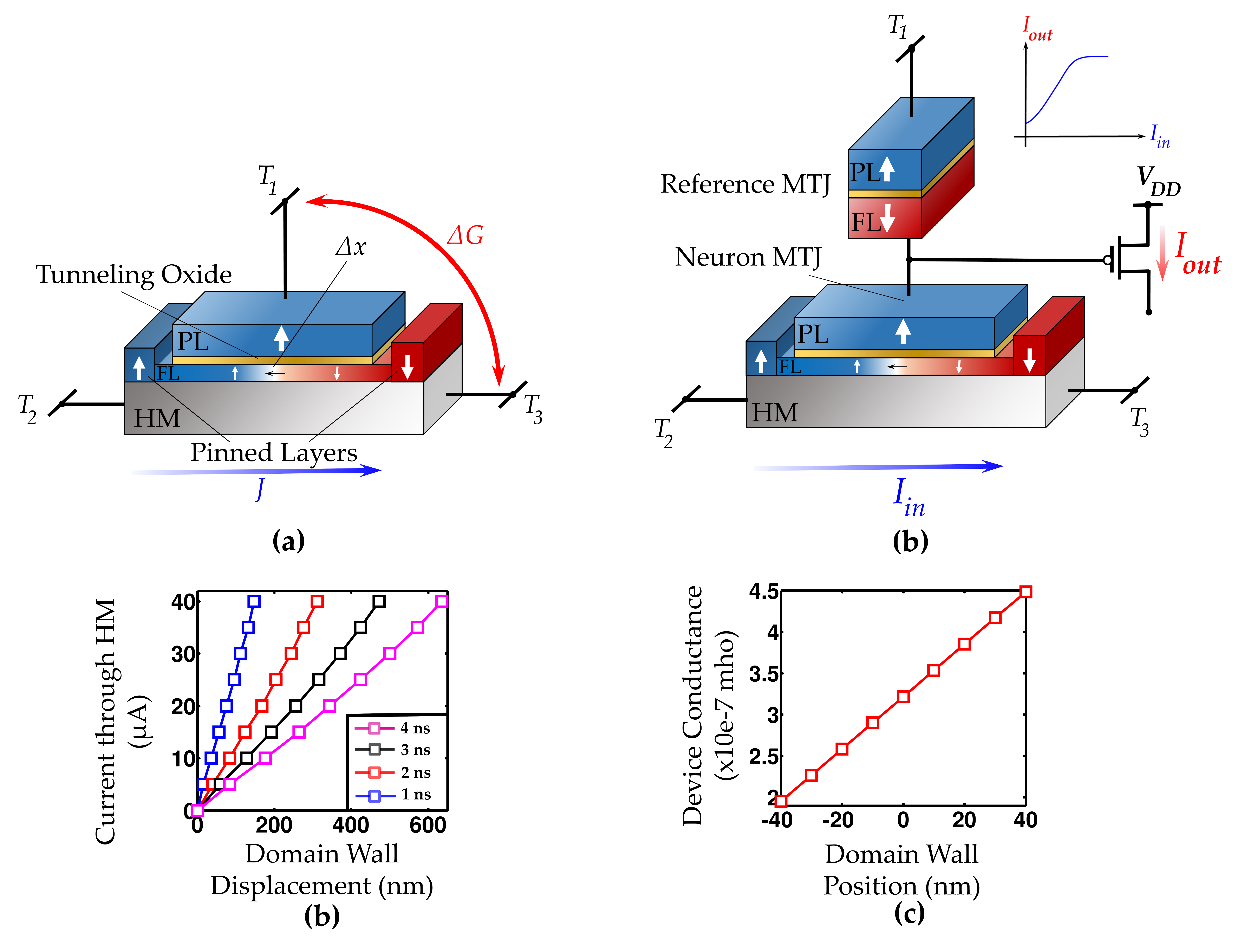}
\caption{(a) DW-MTJ: Magnitude of current flowing through the HM, $J$, causes a proportionate displacement, $\Delta x$, in the DW position, which causes a change, $\Delta G$, in the device conductance between terminals T1 and T3. (b) The same device can be used as a neuron by interfacing with a Reference MTJ. The current provided by the output transistor, $I_{out}$, is a saturated linear function of the input current, $I_{in}$.}
\label{fig3}
\end{figure}

\begin{table}
\label{table1}
\center
\centerline{TABLE I. MTJ Device Simulation Parameters}
\vspace{2mm}
\begin{tabular}{c c}
\hline \hline
\bfseries Parameters & \bfseries Value\\
\hline
Free layer width & $40 nm$\\
Heavy-metal thickness, $t_{HM}$ & $ 2 nm$\\
Saturation magnetization, $M_{S}$ & 1000 $KA/m$ \cite{pai2012spin}\\
Spin-Hall angle, $\theta_{SH}$ & 0.3 \cite{pai2012spin} \\
Energy barrier, $E_{B}$ & 20 $K_{B}T$ \\
Temperature, $T_{K}$ & $300K$ \\
\hline \hline
\end{tabular}\\ 
\end{table}

Considering a worst-case ``reset" current of $140\mu A$ for a duration of $1ns$, the energy consumption involved in using a $20k_{B}T$ barrier magnet (calibrated to experimental measurements reported in \cite{pai2012spin}) as a TRNG is $\sim 57fJ$/bit ($I^2Rt$ energy consumption) \cite{kim2015spin}, which is almost $2\times$ lower than standard MTJ-based TRNG.

\subsection{Domain Wall Motion Based Magnetic Devices - Multi-Level Non-Volatile Memory Design}

The mono-domain magnet discussed above is characterized by only two stable states. For a magnet with elongated shape, multiple domains can be stabilized in the FL, thereby leading to the realization of multiple stable resistive states. Such a domain-wall (DW) MTJ consists of a domain wall separating the two oppositely magnetized regions and the domain wall position is programmed to modulate the MTJ resistance (due to variation in the relative proportion of P and AP domains in the device) \cite{sengupta2016proposal}. 
\begin{figure}[t]
\centering
\includegraphics[width=0.38\textwidth]{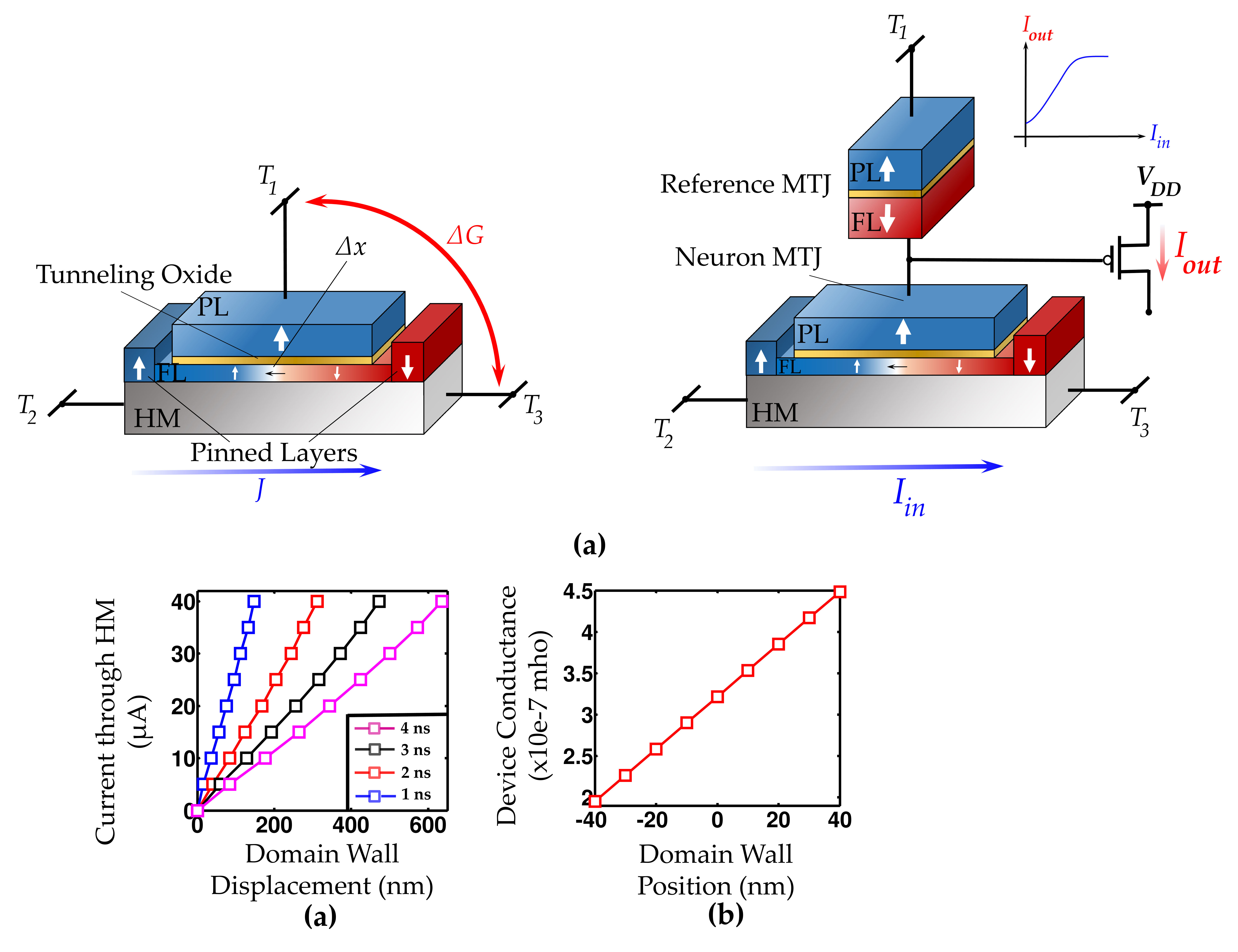}
\caption{Device characteristics are shown for a $20nm$ wide and $0.6 nm$ thick magnet calibrated to experimental measurements \cite{emori2014spin}. The device characteristics illustrate that the programming current magnitude is directly proportional to the amount of conductance change \cite{sengupta2016proposal}. }
\vspace{-5mm}
\label{fig4}
\end{figure}

We consider SHE-based domain wall motion dynamics also in magnet-heavy metal bilayers. In magnetic heterostructures with high perpendicular magnetocrystalline anisotropy, spin-orbit coupling and broken inversion symmetry stabilizes chiral domain walls through Dzyaloshinskii-Moriya interaction (DMI) \cite{emori2013current,emori2014spin}. Such an interfacial DMI at the magnet-heavy metal interface results in the formation of a N\'{e}el domain wall. When an in-plane charge current is injected through the heavy metal, the accumulated spins at the magnet-heavy metal interface results in N\'{e}el domain-wall motion. The device structure is shown in Fig. \ref{fig3}(a), where a current of magnitude, $J$, flowing through the HM layer results in a conductance change, $\Delta G$, between terminals T1 and T3. As shown in Fig. \ref{fig4}(a), for a given programming time duration, the current flowing through the HM underlayer, causes DW displacement proportional to its magnitude. Note that the device characteristics are obtained by performing micromagnetic LLG simulations by dividing the magnet into multiple grids. The domain wall position determines the magnitude of the MTJ conductance. The MTJ conductance varies linearly with the domain wall position since it determines the relative proportion of the area of the Parallel and Anti-Parallel domains of the MTJ (Fig. \ref{fig4}(b)). Since such a device can be programmed to multi-level resistive states and are characterized by low switching current requirements and linear device behavior (device conductance change varies in proportion to magnitude of programming current), they are an ideal fit for implementing crossbar-based ``In-Memory" computing platforms (discussed in next section). We will refer to this device as a DW-MTJ for the remainder of this text. Experimentally, a multi-level DW motion-based resistive device was recently shown to exhibit 15-20 intermediate resistive states \cite{lequeux2016magnetic}.

It is worth noting here that the device structure in Fig. \ref{fig3}(a) can be used as a neuron by interfacing with a Reference MTJ (Fig. \ref{fig3}(b)) \cite{sengupta2016proposal}. The resistive divider can drive a CMOS transistor  where the output drive current would be a linear function of the input current flowing through the heavy metal layer of the device, thereby mimicking the functionality of a Saturated Linear Functionality by ensuring that the transistor operates in the saturation regime \cite{sengupta2016proposal}. The simulation parameters, provided in Table II, were used for the rest of this text for DW-MTJ unless otherwise stated. The parameters were obtained magnetometric measurements of CoFe-Pt nanostrips \cite{emori2014spin}. 

\begin{table}[h]
\label{table2}
\center
\centerline{TABLE II. DW-MTJ Device Simulation Parameters}
\vspace{2mm}
\begin{tabular}{c c}
\hline \hline
\bfseries Parameters & \bfseries Value\\
\hline
Ferromagnet Thickness & $0.6 nm$ \\
Grid Size & $ 4 \times 1 \times 0.6 nm^3$ \\
Heavy Metal Thickness & $ 3 nm$ \\
Domain Wall Width & $ 7.6 nm$ \\
Saturation Magnetization, $M_s$ & 700 $KA/m$ \cite{emori2014spin}\\
Spin-Hall Angle, $\theta_{SH}$ & 0.07 \cite{emori2014spin}\\
Gilbert Damping Factor, $\alpha$ & 0.3 \cite{emori2014spin}\\
Exchange Correlation Constant, $A$ & $1 \times 10^{-11} J/m$ \cite{emori2014spin}\\
Perpendicular Magnetic Anisotropy, $K_{u2}$ & $4.8 \times 10^{5} J/m^{3}$ \cite{emori2014spin}\\
Effective DMI constant, $D$ & $-1.2 \times 10^{-3} J/m^{2}$ \cite{emori2014spin}\\
\hline \hline
\end{tabular}\\ 
\end{table}

\begin{figure*}[t]
\centering
\includegraphics[width=\textwidth]{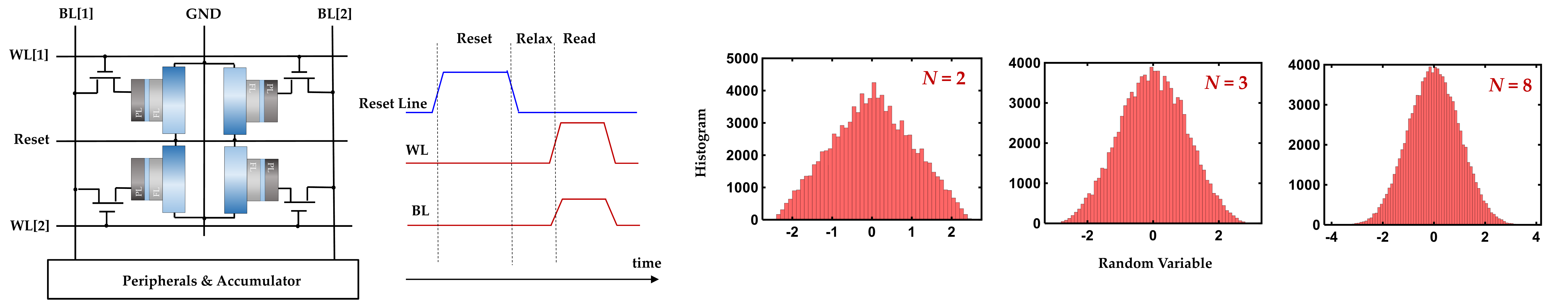}
\caption{(a) Outline of a $2\times 2$ array utilizing spin-based devices interfaced with an accumulator to implement a Gaussian RNG. The probability distributions of random numbers generated from such an array are shown in the extreme right by using a sum of $N$ random variables (rows of the array). We use 8-bit representation and 100,000 samples to plot the distribution. }
\label{fig5}
\end{figure*}
\section{All-Spin Bayesian Neural Networks}
\subsection{Spin-Based Gaussian Random Number Generator} 

Gaussian random number generation task is a hardware-expensive process. CMOS-based designs for Gaussian random number generators would usually require large number of registers, linear feedback circuits, etc. For instance, a recent work for a CMOS-based Gaussian RNG implementation reports $1780$ registers and $528.69mW$ power consumption for a $64$-parallel Gaussian random number generator task \cite{cai2018vibnn}. 

Let us now discuss our proposal of spin-based Gaussian random number generator. In the previous section, we discussed the design of a spintronic TRNG. An array of TRNGs can be used for sampling from a uniform probability distribution. Note that each spin device can be considered to produce a sample from a Bernoulli distribution with probability 0.5. However, reading a particular row of the array provides a sample from a discrete uniform distribution. In order to generate a Gaussian probability distribution from a uniform one, we draw inspiration from statistical Central Limit Theorem, discussed in Box 1. The key result of Central Limit Theorem that we utilize is that the sum of a large number of independent and identically distributed (i.i.d) random variables is approximately Normal. 

\hfill \break  \noindent\fbox{%
		\parbox{.47\textwidth}{%
\textbf{Box 1: Central Limit Theorem}

Let $\{X_1,X_2,...,X_n\}$ be a random sample of $n$ i.i.d random variables drawn from a distribution (which may not be Normal) of mean $\mu$ and variance $\sigma^2$. Then, the probability density function of the sample average, $S_n=\frac{X_1+X_2+...+X_n}{n}$ approaches a Normal distribution with mean $\mu$ and variance $\frac{\sigma^2}{n}$ as $n$ increases. 
}%
	}
\hfill \break 

Our proposed design is illustrated in Fig. \ref{fig5} which depicts a possible array implementation \cite{kim2015spin} of our spin-based TRNGs. Each spin device is interfaced with an access transistor. Rows sharing a Reset-Line can be driven simultaneously. Hence, random numbers can be generated in the entire array in parallel. The timing diagram is shown in Fig. \ref{fig5}. Each row can be read by asserting a particular word-line (WL) and sensing the bit-line voltage (BL). For an $m \times n$ array, each row-read produces an $n$-bit number generated from a uniform probability distribution. By interfacing the array with an accumulator, that averages all the generated random numbers, we are able to produce random numbers drawn from a Normal distribution. Note that the hardware overhead for this process would be high for applications that require precise sampling from Gaussian distributions, since the convergence takes place only for infinite samples. However, for machine learning workloads considered herein, performance of such platforms are usually resilient to approximations in the underlying computations. For instance, Fig. \ref{fig5} shows that even with 8-bit representation and 3 random variables drawn from uniform probability distribution, we are able to achieve an approximate Gaussian distribution. While Gaussian probability distributions are primarily used in such algorithms, non-Gaussian weight distributions can be also designed by using the Gaussian function as a basis. Note that, while Box 1 discussions are equally valid for a CMOS-based TRNG, it will be an order of magnitude more area and power consuming than our proposed spin-based TRNG (as explained in Section III).

\subsection{Dot-Product Operation Between Inputs and Sampled Synaptic Weights}
\begin{figure}[t]
\centering
\includegraphics[width=0.42\textwidth]{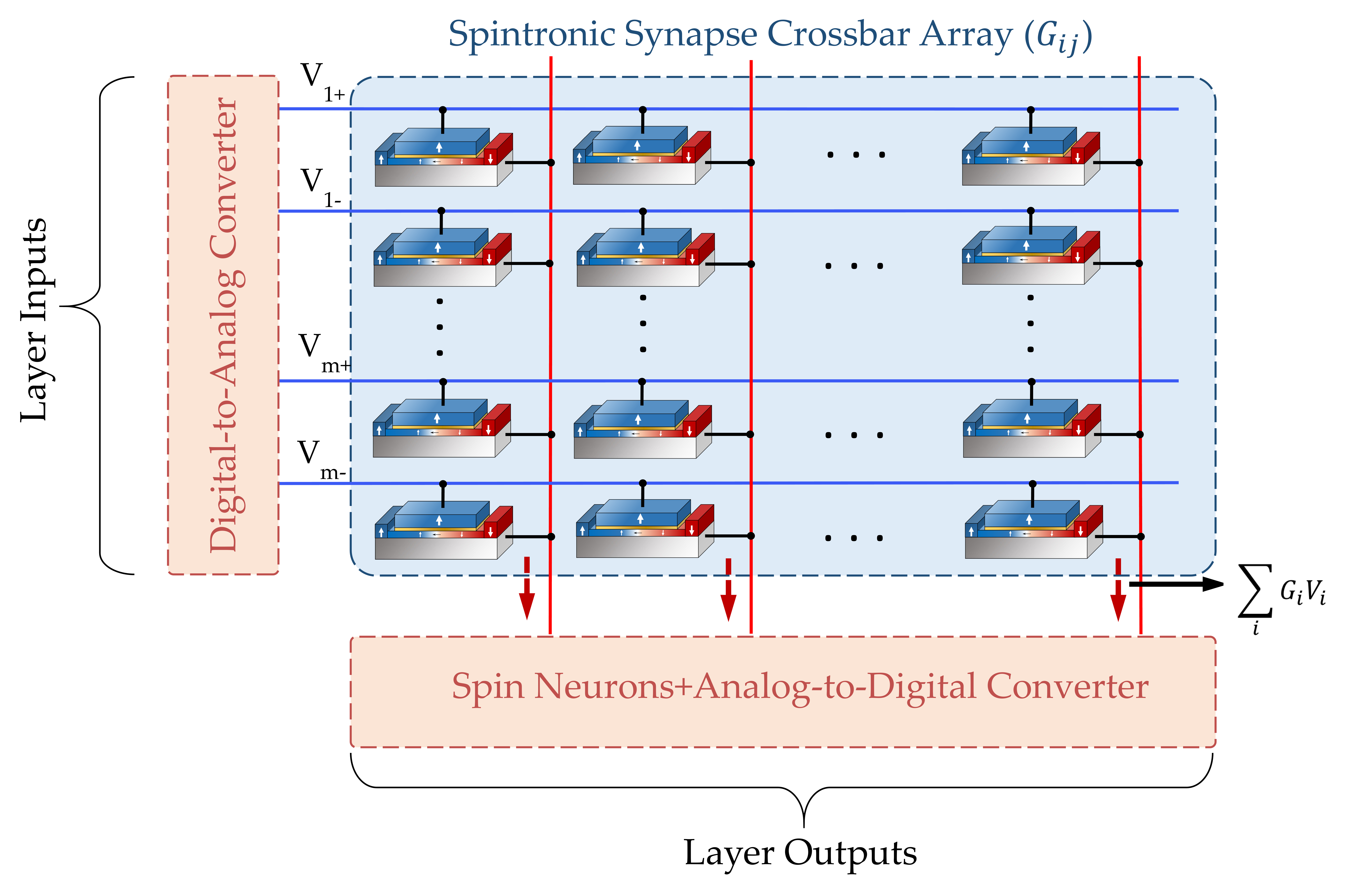}
\caption{ ``In-Memory" computing primitive where an array of spin synapses implement the dot-product kernel.}
\label{fig6}
\end{figure}

Let us first discuss the operation of DW-MTJ enabled spintronic crossbar arrays as an energy-efficient mechanism to realize the dot-product computing kernel. Assuming each synapse to be represented by a DW-MTJ, as shown in Fig. \ref{fig6}, they can be arranged in a crossbar structure. Each row of the array is driven by an analog voltage (output of Digital-to-Analog converters -- DACs) that corresponds to the magnitude of the input. The current flowing through each synapse is scaled by the conductance of the device and due to Kirchoff's law, all these currents get summed up along the column, thereby realizing the dot-product kernel. Note that negative synaptic weights can be also mapped by using two horizontal lines per input (driven by `positive' and `negative' supply voltages). In case a particular synaptic weight is `positive' (`negative'), then the corresponding conductance in the `positive' (`negative') line is set in accordance to the weight. The resultant currents get summed up along the column and pass as the input ``write" current through the spin-neuron. Consecutive ``write" and ``read" cycles of the spin-neurons will implement multiple iterations of the Bayesian network. The analog output current provided by the spin-neuron is then converted to a digital format using the Analog-to-Digital Converters (ADCs). The digital outputs can be latched to provide inputs for the fan-out crossbar arrays. The energy-efficiency of the system stems mainly from two factors: 

\indent {$\bullet$}  The input write resistance of the spintronic neurons are low (being magneto-metallic devices) and they inherently require very low currents for switching. This enables the crossbar arrays of spintronic synapses to be operated at low terminal voltages (typically $100mV$). Further, spintronic neurons are inherently current-driven and thereby do not require costly current to voltage converters, in contrast to CMOS and other emerging technology (Resistive Random Access Memory, Phase Change Memory, among others) based implementations \cite{wijesinghe2018all}.
 
\indent {$\bullet$}  Since, spin devices are inherently non-volatile technologies, the ability to perform the costly Multiply-Accumulate operations in the memory array itself enables us to address the issues of von-Neumann bottleneck. 

\begin{figure}[t]
\centering
\includegraphics[width=0.48\textwidth]{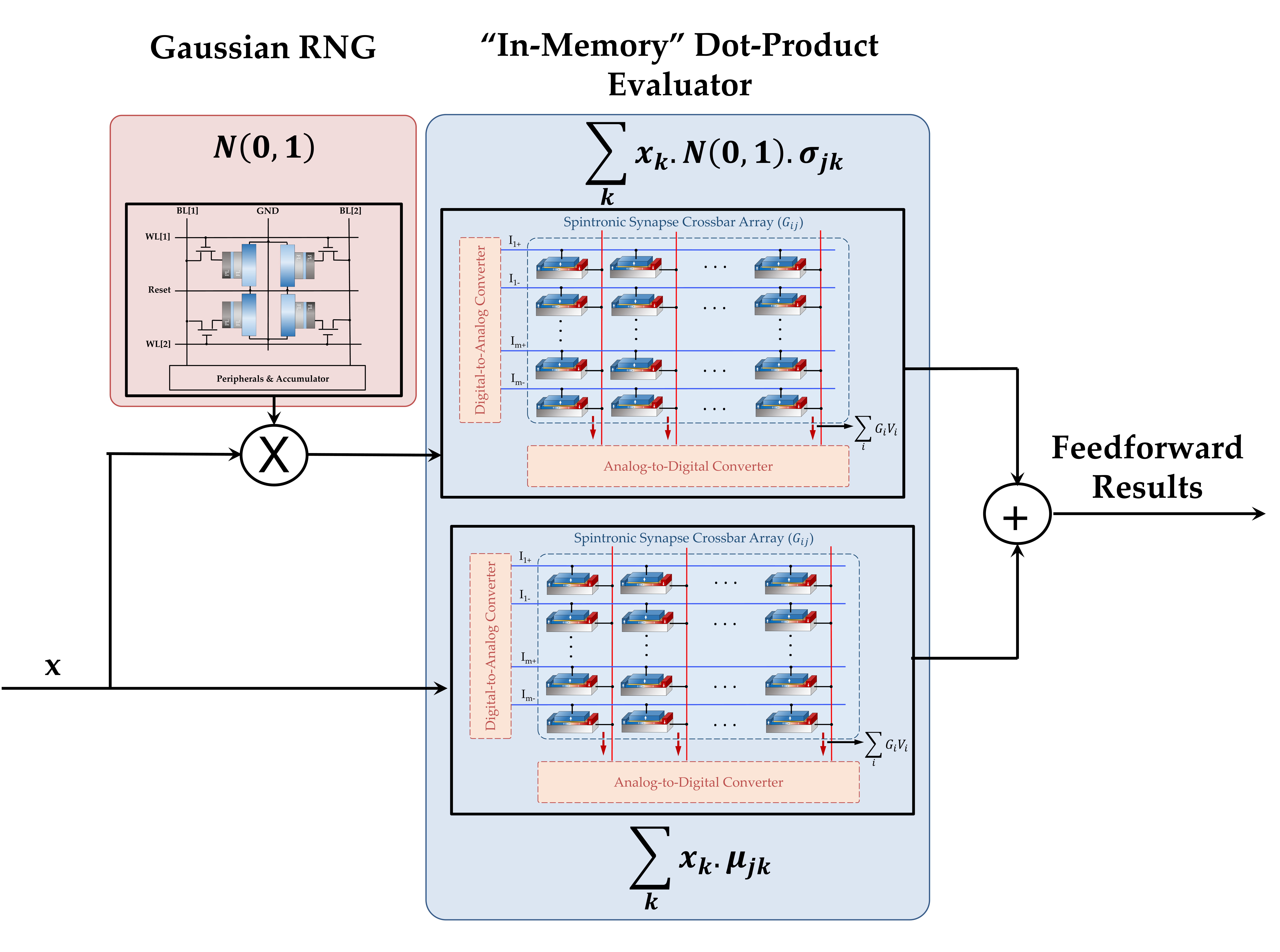}
\caption{All-Spin Bayesian Neural Network Implementation. The two crossbar arrays behave as ``In-Memory" computing kernels whereas the RNG unit provides sampling operation from Gaussian random number generators.}
\label{fig7}
\end{figure}
However, in the context of Bayesian deep networks, even for the inference stage, the synaptic weights are not constant but are updated depending on sampled values from a Gaussian distribution. Assuming we are able to generate samples from a Normal distribution by using the device-circuit primitives proposed earlier, the computations in a Bayesian network can be partitioned in an appropriate fashion such that the benefits of spin-based ``In-Memory" computing can be still utilized. This is explained in Box 2.

Realizing that a Normal distribution with a particular mean and variance is equivalent to a scaled and shifted version of a Normal distribution with zero mean and unit variance, we partition the inference equation as shown in (4). The constant parameters $\mu_{jk}$ and $\sigma_{jk}$ (highlighted in red) represent the mean and variance of the probability distribution of the corresponding synaptic weight and can be therefore implemented by DW-MTJ based memory devices from a hardware implementation perspective. The resultant system (Fig. \ref{fig7}) consists of two crossbar arrays for storing the mean and variance parameters respectively. While the inputs of a particular layer are directly applied to the crossbar array storing the mean values, they are scaled by the random numbers generated from the RNG unit (outputs normalized to provide random numbers with zero mean and unit variance) described previously for the crossbar array storing the variance values. Typical CMOS neuromorphic architectures are characterized by much higher movement of weight data than input data to compute the inference operation \cite{chen2014dadiannao}. Our proposal of computation partition, explained in Box 2, enables us to leverage the ``In-Memory" computing primitives for storing the probability distribution parameters while parallely computing energy-efficient dot-products in-situ between inputs and stochastic weights. It is worth noting here that the crossbar column outputs are computed and read sequentially in order to ensure that the random numbers sampled for the synaptic weights of each column are independent. 

 \hfill \break  \noindent\fbox{%
		\parbox{.475\textwidth}{%
\textbf{Box 2: Computations Involved in Inference Operation}

Once all the \textit{posterior} distributions are learnt ($\mu$ and $\sigma$ parameters of the weight distributions), the network output corresponding to input, $\textbf{x}$, should be obtained by averaging the outputs obtained by sampling from the \textit{posterior} distribution of the weights, $\textbf{W}$ \cite{cai2018vibnn}. The output of the network, $y$, is therefore given by,
\begin{equation}
    y = \mathbb{E}_{P(\textbf{W}|D)}[f(\textbf{x,W})] \approx \mathbb{E}_{q(\textbf{W},\theta)}[f(\textbf{x,W})] \approx \frac{1}{S}\sum_{i=1}^{S}f(\textbf{x,W}^i)
\end{equation}
where, $f(\textbf{x,W})$ is the network mapping for input $\textbf{x}$ and weights, $\textbf{W}$. Using the Variational Inference method, we approximate the weight distribution by Gaussian functions. The approximation is performed over $S$ independent Monte-Carlo samples drawn from the Gaussian distribution, $q(\textbf{W, $\theta$})$.	

Considering just a single layer and neglecting the neural transfer function, $f(\textbf{x,W}^i)$ for the $j$-th neuron can be decomposed into,
\begin{equation}
\begin{aligned}
   f(\textbf{x,W}^i_j) &= \sum_{k}x_k . N(\textcolor{red}{\mu_{jk}},\textcolor{red}{\sigma_{jk}}) \\ &= \sum_{k}x_k . (\textcolor{red}{\mu_{jk}} + \textcolor{red}{\sigma_{jk}}.N(0,1)) \\ &= \sum_{k}x_k . \textcolor{red}{\mu_{jk}} + \sum_{k}x_k   .N(0,1).\textcolor{red}{\sigma_{jk}}
\end{aligned}
\end{equation}
where, $k$ is the dimensionality of the input $\textbf{x}$ and $N(\mu_{jk},\sigma_{jk})$ represent a particular sample drawn from a Normal probability distribution with mean, $\mu_{jk}$, and variance, $\sigma_{jk}$.
}%
	}
\hfill \break 
\vspace{-5mm}

\section{Results and Discussion}
A hybrid device-circuit-algorithm co-simulation framework was developed to evaluate the performance of the proposed All-Spin Bayesian hardware. The magnetization switching characteristics of the mono-domain and multi-domain MTJ was simulated in MuMax3, a GPU accelerated micro-magnetic simulation framework \cite{vansteenkiste2014design}. Non-Equilibrium Green's Function (NEGF) based transport simulation framework \cite{fong2011knack} was used for modelling the MTJ resistance variation with oxide thickness and applied voltage. The obtained device characteristics from MuMax3 and SPICE simulation tools was used in algorithm level simulator, PyTorch, to evaluate the functionality of the circuit. The performance of this design was tested for a standard digit recognition problem on the MNIST dataset \cite{lecun1998gradient}. A two layer fully connected neural network was used, with each hidden layer having 200 neurons. The probability distributions were learnt using the `Bayes by Backprop' algorithm \cite{blundell2015weight}\footnote{The related code can be found at \url{https://github.com/nitarshan/bayes-by-backprop}.}, which learns the optimal Gaussian distribution by minimizing the KL divergence \footnote{The Kullback-Leibler (KL) divergence is a measure of the difference between two probability distributions. In this case, the KL divergence is between the true \textit{posterior}, $P(\textbf{W}|D)$ and the approximated \textit{posterior} $q(\textbf{W},\theta)$. It can be shown that minimization of this difference function can be achieved by using the gradient descent method and iteratively updating the variational parameters, $\mu$ and $\sigma$ \cite{blundell2015weight}.  This is referred to as the `Bayes by Backprop' algorithm.} from the true probability distribution. The prior distribution on the weights used for training was a scaled mixture of two gaussian functions. The network was trained offline to obtain the values of the mean and standard deviation of the probability distributions of the weights. Subsequently they were mapped to the conductances of the DW-MTJ devices. The baseline idealized software network was trained with an accuracy of $98.63\%$ over the training set and $97.51\%$ over the testing set (averaged over 10 sampled networks).  

The device parameters used in this work have been tabulated in the previous section. $20K_{B}T$ barrier height magnet was used in the Gussian RNG unit. We considered 4-bit representation in the DW-MTJ weights and 3-bit discretization in the neuron output. Note that, as explained in the previous section, our neuronal devices mimic a saturating linear functionality and our network was trained with such a transfer function itself. Considering a minimum sensing and programming displacement of $20nm$ for the DW location, we consider our cross-point and neuronal devices to be $320nm$ and $160nm$ in length. From our micromagnetic simulations, we observe the critical current required to switch the neuronal device from one edge to the other is $4\mu A$ for a time duration of $10ns$. The crossbar supply voltage was assumed to be $100mV$ for evaluating the crossbar power consumption. The crossbar resistance ranges (which can be varied by the oxide thickness) were designed to provide the critical current requirement for the spin neurons. We considered $300\%$ TMR in the DW-MTJ conductances of the crossbar array. Considering such device-level behavioral characteristics, non-idealities and constraints, the test accuracy of the network was $96.98\%$ (averaged over 10 samples). Further, non-ideal DW programming can also impact the system accuracy. We performed 5 independent Monte-Carlo runs of the network with $10\%$ variation in each of the programmed crossbar device conductances. The average accuracy degradation was observed to be insignificant - $96.74\%$. 

In order to estimate the system-level energy consumption, we considered the core RNG and crossbar energy consumption along with peripheral circuitry like ADC and DAC\footnote{The energy consumption for the peripheral circuitry were included from typical numbers considered in literature \cite{ankit2019puma,shafiee2016isaac} and can be found at \url{https://github.com/Aayush-Ankit/puma-simulator/blob/training/include/constants.py}.}. We evaluate the energy consumption for a single image inference and a particular network sample. The crossbar read latency was assumed to be $10ns$ (for each column read). During each $10ns$ column read, the power consumption for the DAC and the corresponding crossbar column was considered. Subsequently the neuron device state was read and converted to a digital value using an ADC. The neuron is reset before every operation. For the RNG, DAC and ADC units, we considered 8-bit precision and 3 variables were used for the accumulation process in the Normal distribution sampling. We would like to mention here that we assumed 8-bit precision for the energy calculations in order to achieve a fair comparison with numbers reported in Ref. \cite{cai2018vibnn} for an iso-network CMOS architecture. However, from functional viewpoint, lower bit-precision $\sim 4$ bits was observed to be sufficient. The total energy consumption of our proposed ``All-Spin" network was evaluated to be $790.2nJ$ per classification, which is $24\times$ energy efficient in contrast to the baseline CMOS implementation \cite{cai2018vibnn}. The energy consumption of the RNG unit including peripherals for adding the random numbers generated per row was estimated to be $446.8nJ$. Energy consumption of the crossbar array including DAC, ADC and multiplier peripherals was $343.3nJ$. The system-level energy efficiency stems from both the RNG design and utilization of the ``In-Memory" computing units.

Note that, resistive crossbars are usually characterized by limited fan-in -- much smaller than neuron fan-in in typical deep networks due to non-idealities, parasitics and sneak-paths. Hence, mapping a practically sized network requires mapping synapses of a neuron across multiple crossbars \cite{ankit2019puma,shafiee2016isaac}. Such architectural level innovations can be easily integrated with our current proposal.

\section{Summary}
In summary, we proposed the vision of an ``All-Spin" Bayesian neural processor that has the potential of enabling orders of magnitude hardware efficiency (area, power, energy consumption) in contrast to state-of-the-art CMOS implementations.
Computing frameworks, so far, have mainly segregated deterministic and stochastic computations. Standard deterministic deep learning frameworks enabled by spintronic devices and other post-CMOS technologies have been explored. In such scenarios, device-level non-idealities are usually treated as a disadvantage. More recently, stochasticity inherent in such devices (for instance, probabilistic switching in presence of thermal noise) have been exploited for computing to implement stochastic versions of their deterministic counterparts \cite{sengupta2016probabilistic,srinivasan2016magnetic}. Due to additional information encoding capacity in the switching probability, such devices can be scaled down to single bit instead of multi-bit representations. Device stochasticity has been also used in other unconventional computing platforms like Ising computing, combinatorial optimization problems, among others \cite{roy2018perspective}. Note that prior work on using magnetic devices for Bayesian Inference engines have been proposed \cite{faria2018implementing,shim2017stochastic} which are mainly used for implementing Bayes' rule for simple prediction tasks in directed acyclic graphs and do not have relevance or overlap with Bayesian deep networks. Bayesian deep learning is a unique computing framework that necessitates the merger of both deterministic (dot-product evaluations of sampled weights and inputs) and stochastic computations (sampling weights from probability distributions) - thereby requiring a significant rethinking of the design space across the stack from devices to circuits and algorithms.

\vspace{-2mm}



\end{document}